\documentclass[epj,nopacs]{svjour}
\usepackage{amsfonts,amsmath,amssymb,units,wasysym,epsfig,graphicx,verbatim,color,subfigure,graphicx,bm,mathrsfs,lipsum,hyperref}
\usepackage[utf8]{inputenc}
\usepackage[normalem]{ulem}  

\begin{document}

\title{Radial Oscillations of Neutron Stars with Antikaon Condensate}
\author{Apurba Kheto \inst{1} \and Prasanta Char \inst{2}}
\institute{Government General Degree College, Singur, Jalaghata, Hooghly, West Bengal - 712409, India \and Space  sciences,  Technologies  and  Astrophysics  Research  (STAR)  Institute,  Universit\'{e} de Li\`{e}ge, Bât. B5a, 4000 Li\`{e}ge, Belgium }
\date{}

\abstract{In this article, we study the neutron stars with antikaon condensates and their radial oscillations. We incorporate the antikaons $(K^-,\bar{K}^0)$ within a relativistic mean field theory with density dependent couplings. A transition to the kaon condensed phase is treated as a second order phase transition for different possible antikaon optical potentials. Then, we solve the structure equations of stars and calculate the fundamental and higher order frequencies of their radial pulsations. We study the eigenfunctions of different radial modes for a $2M_\odot$ star with equations of state having different potential depths of antikaons. We also study the large frequency separations for the $2M_\odot$ star. We find distinct features of the appearances of antikaons on the eigenfunctions. The shape of the eigenfunctions corresponding to the radial perturbations is affected due to the appearance of the antikaons, but for the pressure perturbation, the shape remains unchanged.}

\maketitle

\section{Introduction}
\label{intro}
Typical neutron stars (NSs) can have masses $\sim 1- 2.1 M_\odot$ with radii $\sim 10-15$ km. They contain matter at their core with densities at a few times the nuclear saturation densities \cite{Glendenning:1997wn}. These objects offer ideal testbeds for nuclear physics theories beyond nuclear saturation. The structure of the NSs usually depends on the underlying properties of matter. At very high densities, the structure of the ground state of matter may change due to the appearance of strange, non-nucleonic degrees of freedom, favored by the Pauli Exclusion Principle. Consequently, these new degrees of freedom would introduce a softening to the equation of state (EOS) of the NS matter leading to a smaller maximum mass of the star. Unfortunately, the nature of matter at that density regime is poorly understood because such conditions are impossible to create in laboratory experiments. Hence, different types of NS observations are used to constrain the EOS. The first of such constraints came as the precise mass measurements of heavy pulsars. Several pulsars with their masses over $2M_\odot$ have been found \cite{Antoniadis:2013pzd,Arzoumanian:2017puf,Fonseca:2021wxt}. Then in 2017, the extraordinary event GW170817 associated with a binary neutron star merger was observed in gravitational waves (GW), followed by $\gamma$-rays, X-rays, optical and radio waves \cite{TheLIGOScientific:2017qsa,LIGOScientific:2018hze,LIGOScientific:2018cki}. Additionally, the NICER instrument onboard the ISS also reported a couple of informative simultaneous mass-radius measurements \cite{Riley:2019yda,Miller:2019cac,Riley:2021pdl,Miller:2021qha}. These observations led to a plethora of studies of NS matter and a manifold increase in our understanding.  

The study of stellar oscillations has been a very effective tool to understand the interiors of stars. Compact objects like NSs can also produce different types of oscillations due to many different perturbations, e.g., starquakes, pulsar glitches, accretion, tidal perturbations etc. Those oscillations include both radial and different nonradial modes according to the nature of the perturbations and restoring forces. These modes help us to understand the composition of the concerned object. The simplest of these modes is the radial oscillation mode as they do not emit GWs making it the easiest to study. However, it has been shown that they can couple to the nonradial modes thereby modulating and altering the GWs \cite{Passamonti:2005cz,Passamonti:2007tm}. Hence, the signature of radial oscillation, in principle, can be observed in future GW detectors. The possibility of observing radial oscillations has been discussed in literature \cite{Sagun:2020qvc,Sun:2021cez}. It has been proposed that one may detect it through radio waves from pulsars\cite{Boriakoff:1976}, and also in the light curves of short gamma-ray bursts \cite{Chirenti:2019sxw}. For our purpose, we will focus on the radial oscillation spectra of the stable NSs as a tool to study the properties of supranuclear matter. After the seminal works of Chandrasekhar \cite{Chandrasekhar:1964zz,Chandrasekhar:1964zza}, many subsequent studies have been carried out to study radial oscillation for both NSs \cite{1977ApJ...217..799C,1983ApJS...53...93G,1992A&A...260..250V}  and proto-NSs (PNS) \cite{Gondek:1997fd}. Radial oscillation has been used to test different hyposthesis of the dense matter \cite{Sagun:2020qvc,Sun:2021cez,Sen:2022kva,Rather:2023dom}. In this work, we use the same tool to study the effect of phase transition inside dense core of the NSs. Different types of phase transition may occur in the NS matter such as, pions, kaon condensed, deconfined quark matter etc. We concentrate on the Bose-Einstein condensates of antikaons which are particularly favorable strange candidate to appear in the NS core \cite{Schaffner:1995th,Glendenning:1997ak,Schaffner-Bielich:1999fyk}. Kaplan and Nelson first demonstrated the prospect of condensation of $K^-$ in the heavy ion collision \cite{Kaplan:1986yq}.  At higher densities, negatively charges kaons replace the leptons when the lepton chemical potential becomes greater than the in-medium energy of antikaons. The appearance of antikaons has been investigated extensively in the literature in the context of NSs and PNSs \cite{Pal:2000pb,Pons:2000iy,Pons:2000xf,Banik:2000dx,Banik:2001yw,Banik:2002qu,Banik:2014rga,Gupta:2013rba,Char:2014cja,Batra:2017mfv,Thapa:2020usm,Kundu:2022nva}. In most of these studies, the NS matter is described within a relativistic mean field (RMF) theory with the antikaons introduced in the minimal coupling scheme \cite{Glendenning:1997ak}. There are two different types of implementation for the extended RMF models used here. One is with the self-interaction terms added to the original Walecka model \cite{Serot:1984ey} and the other with the density-dependent (DD) couplings of a linear RMF model \cite{Typel:1999yq}. In this work, we will use the DDRMF model, as used in Ref. \cite{Char:2014cja}. The threshold density for antikaons is sensitive to the antikaon optical potential. Deeper potential leads to early appearance of $K^-$ and the consequent softening of the EOS \cite{Banik:2000dx,Banik:2002qu}. One of our goals is to check how deep one can construct such potential with the resulting maximum mass still being consistent with the observations.

There have been many works on the stability of hybrid stars with nonnucleonic phases \cite{Pereira:2017rmp,DiClemente:2020szl,Sun:2021cez}. Most of these studies have considered phase transitions of first order. In such scenarios, the two phases are either separated by a sharp boundary as in Maxwell constructions, or connected via a mixed phase by a Gibbs construction. On the other hand, we are interested to study what happens if the Bose condensates appear within a second order phase transition where the two phases are connected continuously without any mixed phase. Since we want to see the effects of antikaons, we have deliberately excluded the hyperons that are also likely to appear inside the NSs. The paper is organized as follows. In section \ref{matter}, we have briefly summarized the formalism to describe the antikaon condensates in the NS matter. In section \ref{structure}, we have discussed the necessary equations of NS structure and radial oscillations. Then, we have presented our results in section \ref{results}. Finally, we summarize our conclusions in section \ref{summary}.

\section{Description of antikaon condensed matter}
\label{matter}
To describe the NS matter, we have used the DDRMF model, denoted by the Lagrangian density of the form
\begin{eqnarray}
\label{eq_lag_b}
{\cal L}_B &=& \sum_N \bar\psi_{B}(i\gamma_\mu{\partial^\mu} - m_B
+ g_{\sigma B} \sigma - g_{\omega B} \gamma_\mu \omega^\mu 
 \nonumber\\
&& -  g_{\rho B} 
\gamma_\mu{\mbox{\boldmath $\tau$}}_B \cdot 
{\mbox{\boldmath $\rho$}}^\mu  )\psi_B + \frac{1}{2}\left( \partial_\mu \sigma\partial^\mu \sigma
- m_\sigma^2 \sigma^2\right)
 \nonumber\\
&& -\frac{1}{4} \omega_{\mu\nu}\omega^{\mu\nu} +\frac{1}{2}m_\omega^2 \omega_\mu \omega^\mu  \nonumber\\
&& - \frac{1}{4}{\mbox {\boldmath $\rho$}}_{\mu\nu} \cdot
{\mbox {\boldmath $\rho$}}^{\mu\nu}
+ \frac{1}{2}m_\rho^2 {\mbox {\boldmath $\rho$}}_\mu \cdot
{\mbox {\boldmath $\rho$}}^\mu.
\end{eqnarray} 

The meson-baryon couplings are the functions of total baryon density. The functional form for the density-dependence and the associated parameters are taken from Typel et al. \cite{Typel:2009sy}.  While computing the mean field equations, the variation of ${\cal L}_B$ results in a rearrangement term due to the density dependence of the couplings. The effective baryon mass is defined as $m_B^*=m_B-g_{\sigma B}\sigma$, with $m_B$ as the vacuum rest mass of baryon B. The vector self-energy is given by, $\Sigma_{B}=\Sigma^{(0)}_{B}+\Sigma^{(r)}_B$. The first term in $\Sigma_{B}$ consists of the non-vanishing components of the vector mesons i.e. $\Sigma^{(0)}_{B}=g_{\omega B}\omega_0 + g_{\rho B} \tau_{3B} \rho_{03}$. The second term is the rearrangement term that takes the form, 

\begin{equation}\label{eq_rear}
\Sigma^{(r)}_B=\sum_B[-g_{\sigma B}'  
\sigma n^{s}_B + g_{\omega B}' \omega_0 
n_B
+ g_{\rho B}'\tau_{3B} \rho_{03} n_B ],
\end{equation}
where $g_{\alpha B}'=\frac {\partial g_{\alpha B}} {\partial \rho_B}$, $\alpha= \sigma,~ \omega, ~\rho$, and $\tau_{3B}$ is the isospin projection values of $B=n,p$. Inside the NSs, the baryons and leptons are in chemical equilibrium governed by the general equilibrium condition, $\mu_i = b_i \mu_n - q_i \mu_e ~$. Here, $b_i$ is the baryon number and $q_i$, the charge of $i$-th baryon, $\mu_n$ and $\mu_e$ are the chemical potentials of neutrons and electrons, respectively. 

Now, we discuss the antikaon condensed phase composed of the nucleons, and the antikaon isospin doublet with electron and muons in the background. The interaction among the nucleons in this phase is again described by the Lagrangian density of Eq. (\ref{eq_lag_b}). We choose the antikaon-baryon interaction on the same footing as the baryon-baryon interaction. The Lagrangian density for antikaons in the minimal coupling scheme is given by, 
\begin{equation}
{\cal L}_K = D^*_\mu{\bar K} D^\mu K - m_K^{* 2} {\bar K} K ~,
\end{equation}
where the covariant derivative is $D_\mu = \partial_\mu + ig_{\omega K}{\omega_\mu} + i g_{\rho K} {\boldsymbol \tau}_K \cdot {\boldsymbol \rho}_\mu + i g_{\phi K} {\phi_\mu}$, and the effective mass of (anti)kaons is given by $m_K^* = m_K - g_{\sigma K} \sigma $, where $m_K$ is the bare kaon mass. Here, we have included the $\phi$-meson which is widely used as strangeness mediating interaction among hyperons \cite{Ellis:1990qq,Schaffner:1995th}. The isospin doublet for kaons is denoted by $K\equiv (K^+, K^0)$ and that for antikaons is $\bar K\equiv (K^-, \bar K^0)$. For s-wave (${\bf p}=0$) condensation, the in-medium energies of $\bar K\equiv (K^-, \bar K^0)$ are given by,
\begin{equation} \label{ch1.kom}
\omega_{K^-,\: \bar K^0} = m_K^* - g_{\omega K} \omega_0 - g_{\phi K} \phi_0
\mp  g_{\rho K} \rho_{03}.
\end{equation}
For the $s$-wave ({\bf k}=0) ${\bar K}$ condensation at zero temperature, the scalar and vector densities of antikaons are same and those are given by,
\begin{equation}
n_{K^-,\: \bar K^0}=2\left( \omega_{K^-, \bar K^0} + g_{\omega K} \omega_0
+ g_{\phi K} \phi_0 \pm  g_{\rho K} \rho_{03} \right) {\bar K} K~.
\end{equation} 
The requirement of chemical equilibrium fixes the threshold condition of antikaon condensations in NS matter.
\begin{eqnarray}
\mu_n - \mu_p &=& \mu_{K^-} = \mu_e ~, \\
\mu_{\bar K^0} &=& 0 ~,
\end{eqnarray}
where $\mu_{K^-}$ and $\mu_{\bar K^0}$ are the chemical potentials of $K^-$ and $\bar K^0$, respectively. In this exercise, we do not consider the kaon-meson couplings density dependent. The scalar coupling constant is calculated from the antikaon optical potential at the symmetric nuclear matter at saturation \cite{Schaffner:1995th,Banik:2001yw,Char:2014cja}. We have used the vector meson couplings from quark model and isospin counting rule \cite{Schaffner:1995th,Glendenning:1997ak}. 

\section{Stellar structure and radial oscillations}
\label{structure}

For a static and spherically symmetric spacetime relevant to a compact star is described by the metric,
\begin{equation}
    ds^2 = - e^{\nu(r)}dt^2 + e^{\lambda(r)}dr^2 + r^2(d\theta^2 + sin^2d\phi^2),
\end{equation}
where, the $\nu(r)$ and $\lambda(r)$ are the metric functions given by the solutions of the Einstein equations. The structure of a stable NS at hydrostatic equilibrium is given by the Tolman-Oppenheimer-Volkoff (TOV) equations,
\begin{eqnarray}
    \frac{dP}{dr}  &=&  - \frac{(P+\epsilon)(m+4\pi r^3 P)}{r(r-2m)},  \\
    \frac{dm}{dr}  &=& 4 \pi r^2 \epsilon , \\
   \frac{d\nu}{dr} &=& - \frac{2}{P+\epsilon} \frac{dP}{dr} ~,
\end{eqnarray}
where, the $\epsilon$, $P$, and $m$ are the energy density, pressure, and the enclosed mass respectively. The radius of the star, $R$ is defined as the surface where $P(R)=0$ and then the mass of the star becomes, $M=m(R)$. 

We are interested in those solutions of the TOV equations that are stable against radial perturbations. The equations governing the radial oscillations were first derived by Chandrasekhar in the Sturm-Liouville form \cite{Chandrasekhar:1964zza}. For numerical purposes, Vaeth and Chanmugam rewrote them in terms of two first order equations for the quantities $\Delta r/r$ and $\Delta P/P$ \cite{1977ApJ...217..799C,1992A&A...260..250V}. Gondek et al. \cite{Gondek:1997fd} also obtained a different set of two first order equations for radial displacement, $\xi = \Delta r/r$ and Lagrangian perturbation of pressure $\Delta P$. The advantage of the latter approach  is that there is no singularity at the stellar surface. Hence, we solve the following equations by Gondek et al. \cite{Gondek:1997fd} to find the oscillation frequencies. 
\begin{eqnarray}
  \frac{d \xi}{dr}&=&
- \frac{1}{r}\left(3\xi+ \frac{\Delta P}{\Gamma P}\right)-
\frac{dP}{dr}\frac{\xi}{(P+\epsilon)}~, \label{osc1}\\
\frac{d \Delta P}{dr}&=&
\xi\left\{ 
\omega^2
e^{\lambda-\nu}\left( P+\epsilon \right)r
-4\frac{dP}{dr}\right\}\nonumber\\
&+&\xi\left\{
 \left(\frac{dP}{dr}\right)^2 
\frac{r}{(P+\epsilon)}
- 8\pi e^\lambda (P+\epsilon)Pr\right\}\nonumber\\
&+&{\Delta P} \left\{\frac{dP}{dr} \frac{1}{(P+\epsilon)}- 4\pi
(P+\epsilon)r e^\lambda 
\right\}~, \label{osc2}
\end{eqnarray}
where $\Gamma$ is a adiabatic index, $\omega$ is the eigenfrequency. The eigenfunctions, $\xi$ and $\Delta P$ are assumed to have a harmonic time dependence $\propto e^{i\omega t}$. 
To solve these equations, we require two boundary conditions which we get from the requirements of regularity at the center and the vanishing pressure at the surface, respectively. As $r \rightarrow 0$, it translates as the coefficient of the $1/r$ term of Eqn. \ref{osc1} going to zero as well. Hence,
\begin{equation}
    (\Delta P)_{r \rightarrow 0} = -3(\xi \Gamma P)_{r \rightarrow 0}.
\end{equation}
Then, the eigenfunction can be normalized to get $\xi(0)=1$. For the surface, the boundary condition is expressed as the Lagrangian perturbation of pressure becomes zero. 
\begin{equation}
    (\Delta P)_{r=R} = 0.
\end{equation}
The solutions of the above equations provide discreet eigenvalues $\omega_i^2$ for a given static NS model with the stability criteria being $\omega_i^2 \gtrsim 0$.

In a Sturm-Liouville boundary value problem, the number of zeros in the eigenfunctions corresponds to the overtone number $ n$, for example, the first excited mode, corresponding to $n = 1$, has only one zero, the second excited mode, corresponding to $n = 2$, has two zeros, whereas the fundamental mode, corresponding to $n = 0$, has none. The fundamental mode is also known as the f-mode, while the rest of the modes with $n =1;2; 3; ...$ are the so-called p-modes (pressure modes or acoustic modes)\cite{Sagun:2020qvc}.

\section{Results and Discussions}
\label{results}

Now, we discuss the numerical results of the radial oscillation of NSs with antikaon condensates ($K^-$,${\bar {K^0}} $). We have generally followed the calculations in Ref. \cite{Char:2014cja} in constructing the EOSs. We consider antikaons in the nucleonic system consisting of neutrons, protons, electrons and muons. We have used the DD2 parameter set for the density-dependent meson-baryon couplings in our calculation. The symmetry energy and its slope parameter for the DD2 model is $31.67$ MeV and $55.04$ MeV, respectively. These values are consistent with the latest constraints from chiral effective field theory calculations \cite{Lim:2023dbk}.  For the low-density part of EOS concerning the crust, we have used the EOS provided by Baym, Pethick, and Sutherland \cite{Baym:1971pw}. The effect of $K^-$ condensates in NS matter is to maintain charge neutrality by displacing electrons and softening the EOS. This results in a decrease in the maximum mass of the NS. We choose a set of values of antikaon optical potential ( $U_{\bar K}$) from $-100$ to $-180$ MeV. The maximum mass and radius of the stars for different EOSs using different antikaon optical potentials are shown in Table \ref{tab1}.

\begin{table}
    \centering
    \begin{tabular}{ccc}
    \hline\hline
     
      $U_{\bar K} (MeV)$& $M_{max}$  ($M_{\odot}$) & $R_{max}$ (km)  \\ \hline
      - 100  & 2.298 & 12.15 \\
      - 110  & 2.271 & 12.10  \\
      - 120  & 2.241 & 12.06   \\
      - 130  & 2.206 & 12.04   \\
      - 140  & 2.163 & 12.02 \\
      - 150  & 2.113 & 11.97 \\
      - 160  & 2.053 & 11.89 \\
      - 170  & 1.982 & 11.74\\
      - 180  & 1.900 & 11.46 \\
      np $ e^- \mu^- $ &  2.417 & 11.87
    \end{tabular}
    \caption{Maximum mass and radius of NSs for different EOS with different antikaon potentials.}
    \label{tab1}
\end{table}

\begin{figure}
    \centering
    \includegraphics[scale=.35]{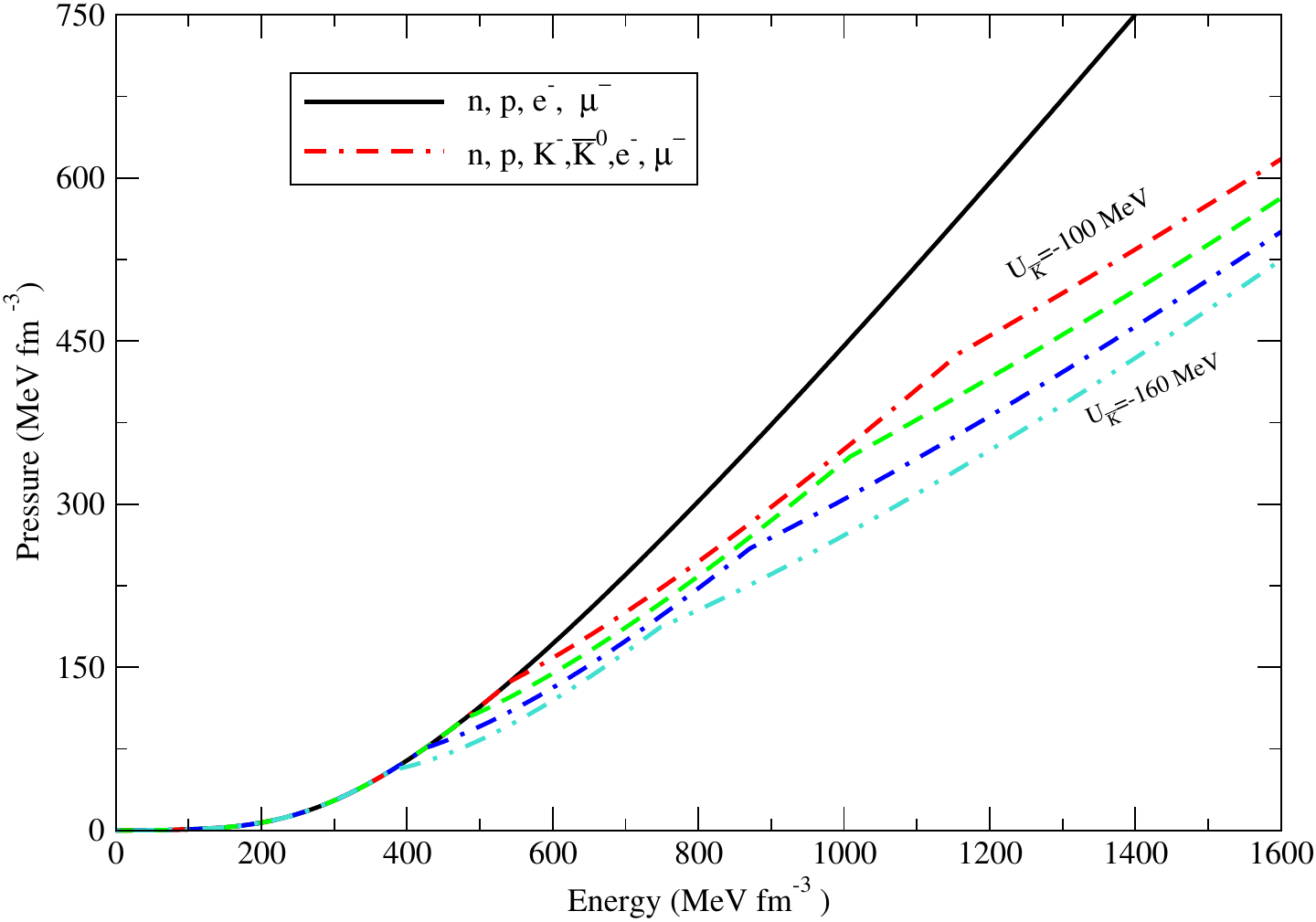}
    \caption{Pressure is shown as a function of energy density. The solid line corresponds to the  nucleon-only matter and the dashed lines imply that antikaon condensates with different optical potentials from $-100$ to $-160$ MeV.}
    \label{fig:eos}
\end{figure}

\begin{figure}
    \centering
    \includegraphics[scale=.35]{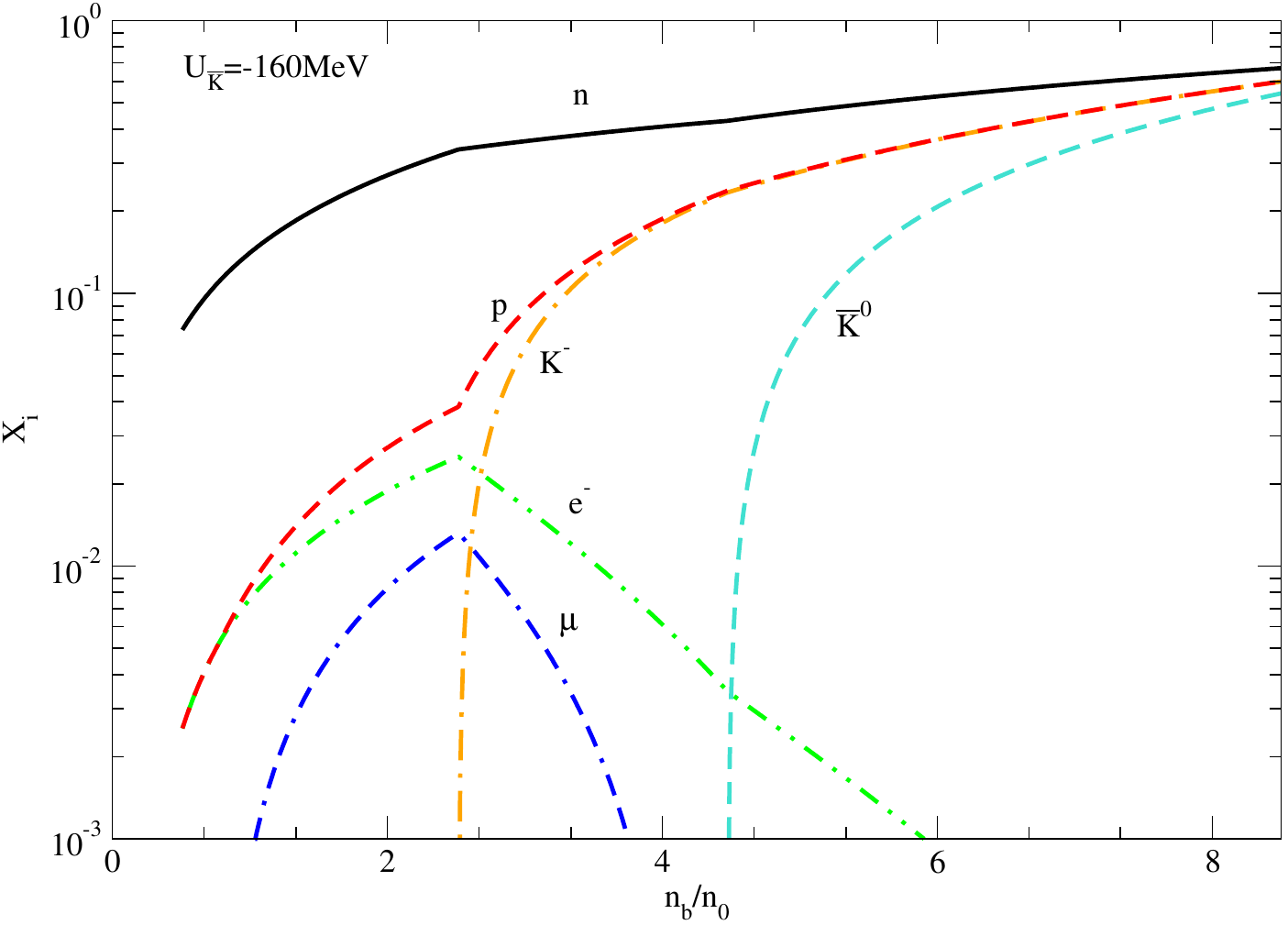}
    \caption{Particle fraction is shown as a function of normalized baryon density for $U_{\bar K} = -160$ MeV. }
    \label{fig:frac}
\end{figure}

\begin{figure}
    \centering
    \includegraphics[scale=.40]{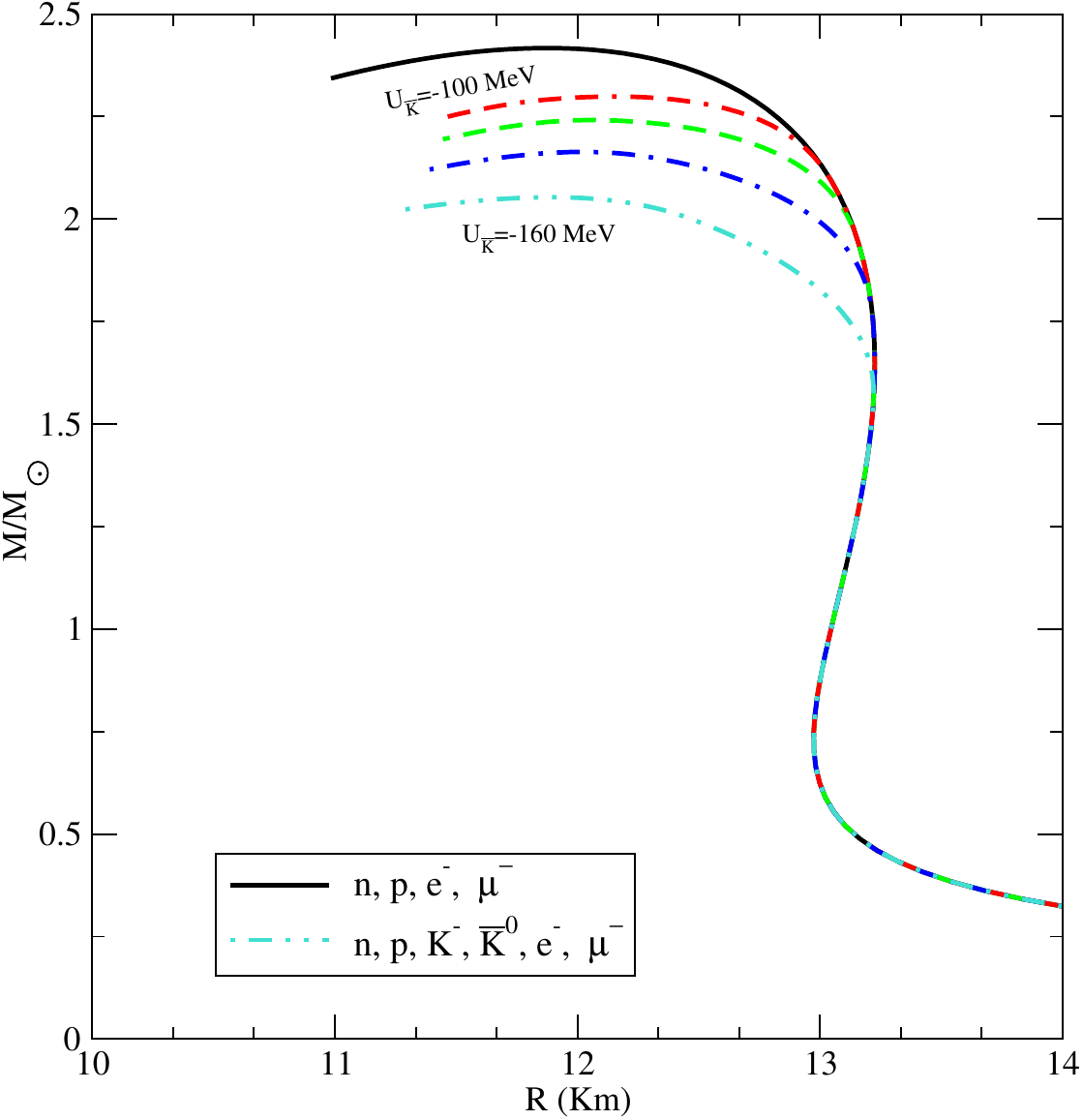}
    \caption{NS sequence is plotted with the radius of the star corresponding to the EOSs of Fig. \ref{fig:eos}.}
    \label{fig:mr}
\end{figure}

The maximum mass constraint (above $2 M_{\odot}$) excludes the antikaon optical potential deeper than $-170$  MeV. So for the rest of this section, we use EOS of nucleon-only matter and EOSs  with antikaon condensates having  optical potentials between $-100$ to $-160$ MeV. For all these potentials, the antikaons appear in the system as a second order phase transition. This is clear from the Fig. \ref{fig:eos} where we  have plotted the EOSs as the variation of pressure with energy density. We do not see any decrease in the energy density when the antikaons start appearing. The solid line corresponds to the nucleon-only matter consisting of neutrons, protons, electrons, and muons whereas the other dashed lines correspond to the matter including $K^-$ and $\bar K^0$ with antikaon optical potential ($U_{\bar K}$) from $-100$ to $-160$ MeV. For the value of $U_{\bar K}=-160$ MeV the EoS is softest. The first kinks indicate the appearance of $K^-$ and the other kinks in the higher energy range indicates the appearance of $\bar {K^0}$ in dashed lines. For our deepest potential $U_{\bar K}= -160$ MeV, we have also shown the particle fractions in Fig. \ref{fig:frac}. We see that the $K^-$ appears as low as $\sim 2.52 n_0$ because of the very deep optical potential. This results in a decrease in the population of leptons and completely replacing them at higher densities. The mass-radius sequences of the NSs corresponding to the EOSs in Fig. \ref{fig:eos} are shown in Fig. \ref{fig:mr}.  The solid line represents the mass of the sequence of the stars using nucleon-only EOS whereas the dashed lines represent the masses of the star sequences  using the EoS with antikaon condensates having optical potential $-100$ to $-160$ MeV.  We notice that in all the cases the values of the maximum mass exceed the observational benchmark  $2 M_\odot$ as shown in Fig. \ref{fig:mr}. We used these five EOSs to study NS radial oscillation.

\begin{figure}
    \centering
    \includegraphics[scale=.35]{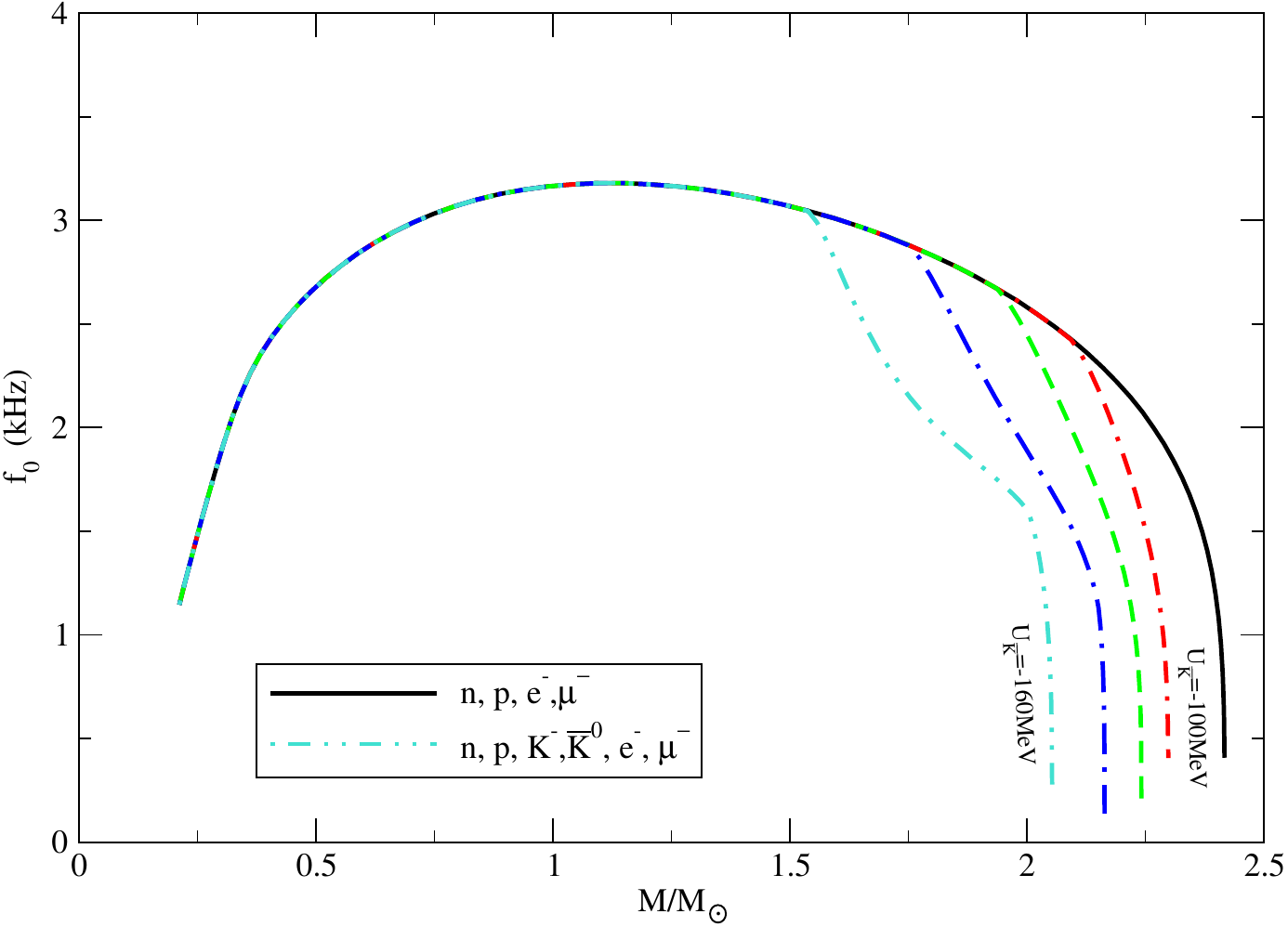}
    \caption{Fundamental frequencies are plotted as a function of the mass of NS sequences corresponding to the EOSs of Fig. \ref{fig:eos}.}
    \label{fig:f0}
\end{figure}

\begin{figure}
    \centering
    \includegraphics[scale=.35]{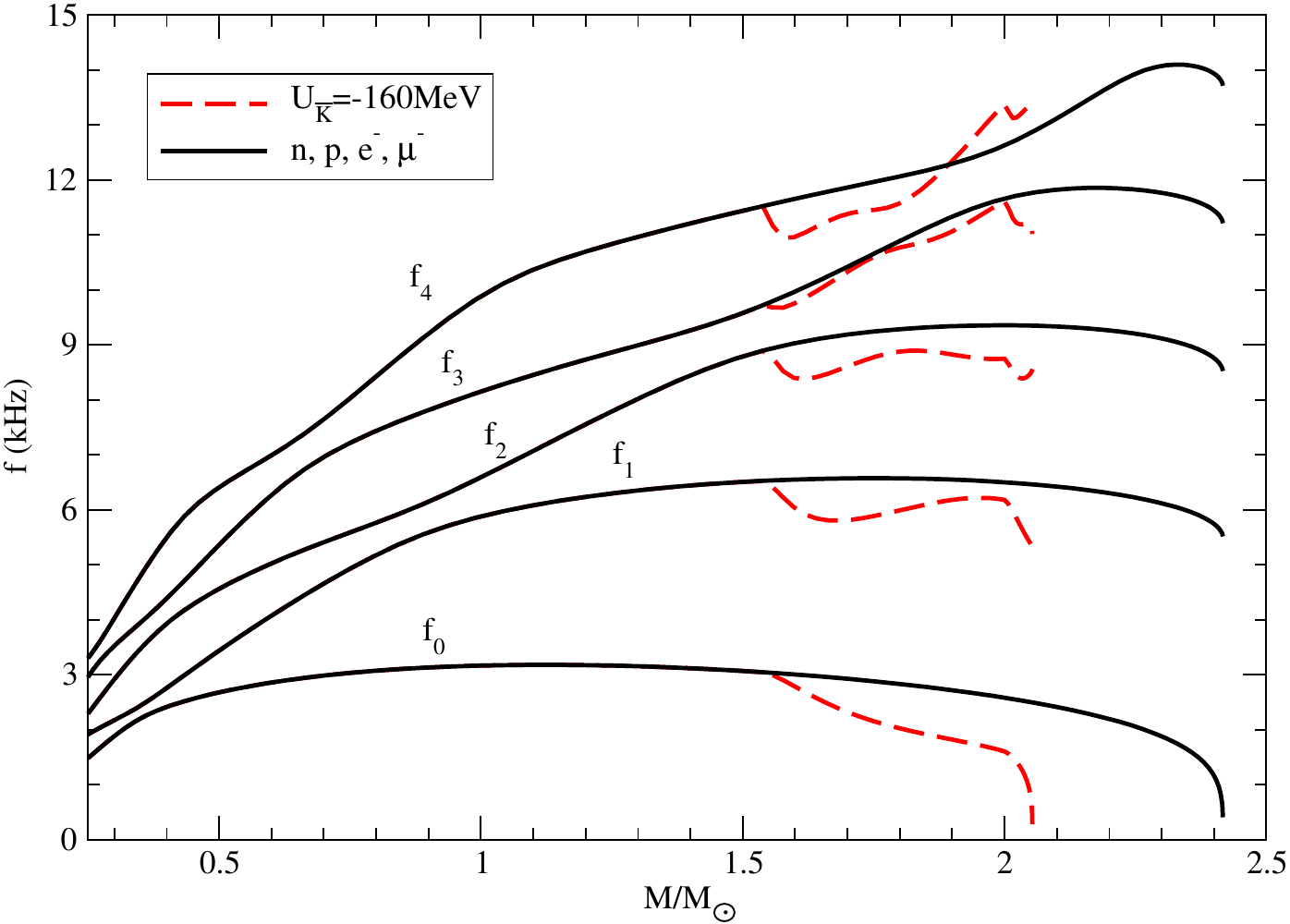}
    \caption{Fundamental and higher order frequencies are plotted as a function of the mass of NS sequences. The solid line corresponds to the  nucleon-only matter and the dashed line implies that antikaon condensates with  $U_{\bar K}=-160$ MeV. }
    \label{fig:ff}
\end{figure}

In Fig. \ref{fig:f0}, we have presented the plot of the fundamental frequencies as a function of the mass of the stellar sequences with the EOSs of Fig. \ref{fig:eos}. The fundamental frequencies increase initially as  the mass of the star increases and then decrease after $\sim 1 M_{\odot}$. As the mass-radius sequence attains its maximum value, we notice a sharp fall as expected because of the unstable region. With the dashed lines, we have the fundamental frequencies of the sequences with antikaon condensates with four different optical potentials ranging from $-100$ to $-160$ MeV. We can clearly see in this figure that fundamental frequencies decrease rapidly as soon as antikaon is produced. For 
$U_{\bar K}=-100 MeV$ the change is small from nucleon-only matter but for $U_{\bar K}=-160 MeV$, the change is quite large. Two distinct kinks appear in the dashed line indicating the appearances of $K^-$ and  $\bar {K^0}$.  In Fig. \ref{fig:ff}, we have shown the fundamental frequencies as well as a few higher-order frequencies as a function of the mass of the NS sequence. Here we see that higher-order frequencies behave differently compared to the fundamental frequencies due to the onset of antikaon condensates and also the number of kinks increases in higher-order modes. This is apparently due to the avoided crossings \cite{Gondek:1999ad,Kokkotas:2000up}. In our case, the kinks are more prominent because of the significant amounts of antikaons in the system for higher mass stars. In $f_4$, the frequency of antikaon condensed matter exceeds the frequency of normal nuclear matter. So the nature of fundamental  and  overtone frequencies of the NS sequence is completely changed when antikaon condensates appear in the high-density nuclear matter. Hence, the radial oscillation measurements of stars can provide deeper knowledge of the microphysics of the NS interior.

\begin{table}
    \centering
    \caption{ Frequencies $\nu_n$ in kHz for the radial modes of a $2M_\odot$ star with three different EOSs, with n denoting the order of the modes. }
    \begin{tabular}{cccc}
    \hline\hline
     
      n & $n p e^- \mu^-$  &$U_{\bar K}=-120 MeV$ & $U_{\bar K}=-160 MeV$   \\ \hline
      0  & 2.57528 &2.42170 &1.55987 \\
      1  & 6.49845 &6.14911 &6.09897 \\
      2  & 9.35687 &8.85951 &8.64465  \\
      3  & 11.6733 &11.2635 &11.4720  \\
      4  & 12.6482 &12.4596 &13.2346\\
      5  & 14.4784 &14.0193 &14.1355\\
      6  & 16.7281 &16.2013 &16.1408\\
      7  & 18.9109 &18.4605 &18.2671\\
      8  & 20.0377 &19.6796 &20.3385\\
      9 & 22.2253 &21.4041 &21.8449   \\
      10 & 24.1869 &23.4365 &23.6887   \\
      \hline\hline
      
    \end{tabular}
    
    \label{tab2}
\end{table}

\begin{figure}
    \centering
    \includegraphics[scale=.35]{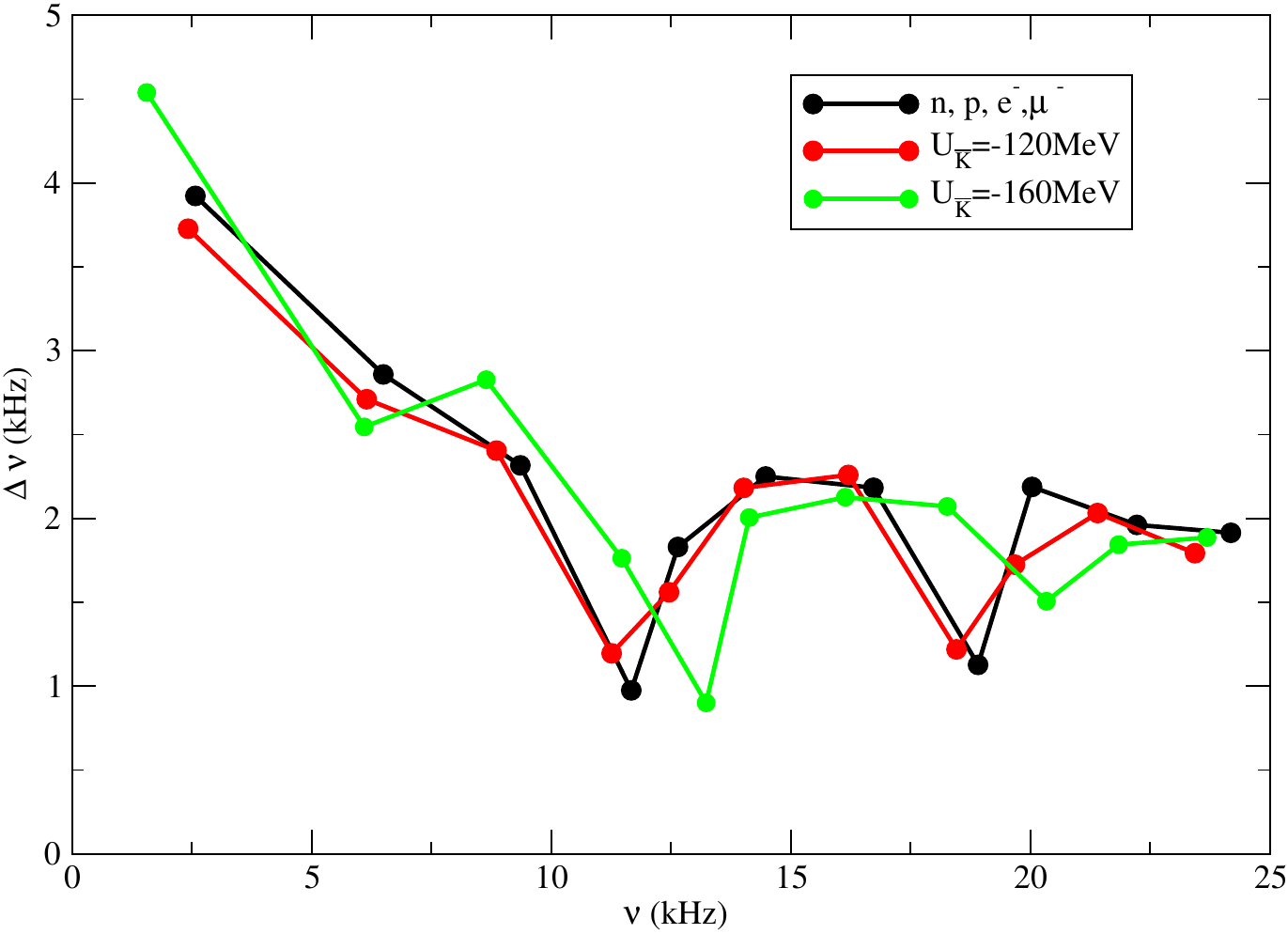}
    \caption{Large frequency differences are shown as a function of frequency (both in kHz) for three stars of the same mass ($2 M_{\odot}$).  The black  line corresponds to the  nucleon-only matter, the red and green lines imply the antikaon condensates with  the optical potentials  of $-120$ MeV and $-160$ MeV, respectively.  }
    \label{fig:fdif}
\end{figure}

Table ~\ref{tab2} shows the frequencies of the first $11$ radial modes for the three different EOSs with the same mass. In asteroseismology, a difference between consecutive modes  which is called the large separation i.e., $\Delta \nu_n=\nu_{n+1}-\nu_n$, is widely used to interpret star properties \cite{Sagun:2020qvc,Sun:2021cez,Rather:2023dom}. In Fig. \ref{fig:fdif}, we show the comparison of the $\Delta \nu_n$  as a function of frequency (both in kHz) for three stars of the same mass ($2 M_{\odot}$). The frequencies differences of the neutron star made of nucleon-only matter are shown in black; the large separation of NS made of  antikaon condenses with an optical potential of $-120$ MeV is shown in red; and antikaon condenses with an optical potential of $-160$ MeV is shown in green. Here we observe the erratic fluctuations of $\Delta \nu_n$ due to the presence of crust in our EOSs. The variation in frequency, $\nu_n$, for a given EoS without crust is smooth, as discussed in Refs.\cite{Sagun:2020qvc,Panotopoulos:2018ipq}.

\begin{figure}
    \centering
    \includegraphics[scale=.30]{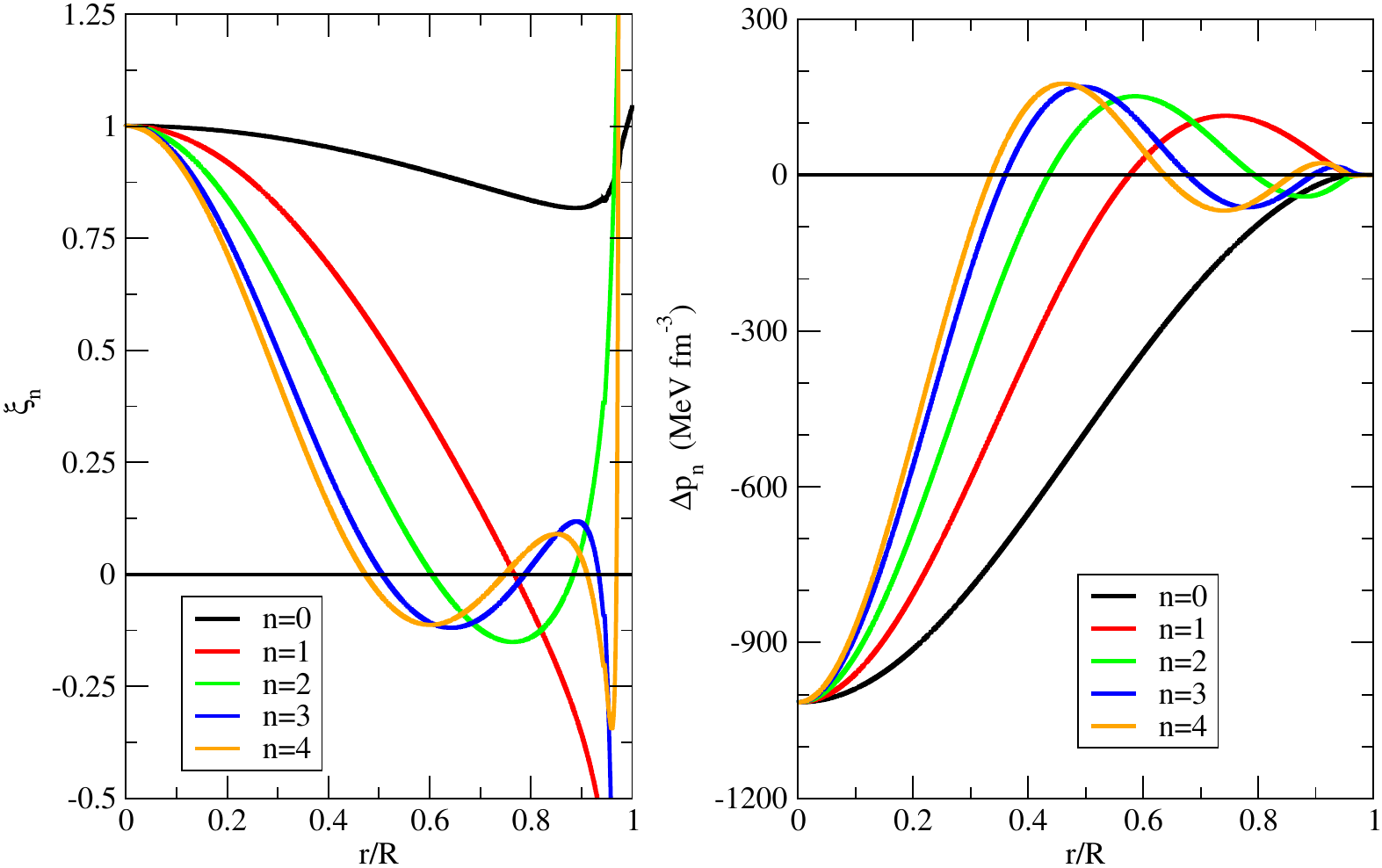}
    \caption{Left Panel: First eigenfunctions are  plotted as a function of the radial distance of a NS of mass ($2 M_{\odot}$) for the nucleon-only matter.  Right Panel: Second eigenfunctions are  plotted as a function of the radial distance of a NS of mass ($2 M_{\odot}$) for the nucleon-only matter.  }
    \label{fig:xinpe}
\end{figure}

\begin{figure}
    \centering
    \includegraphics[scale=.32]{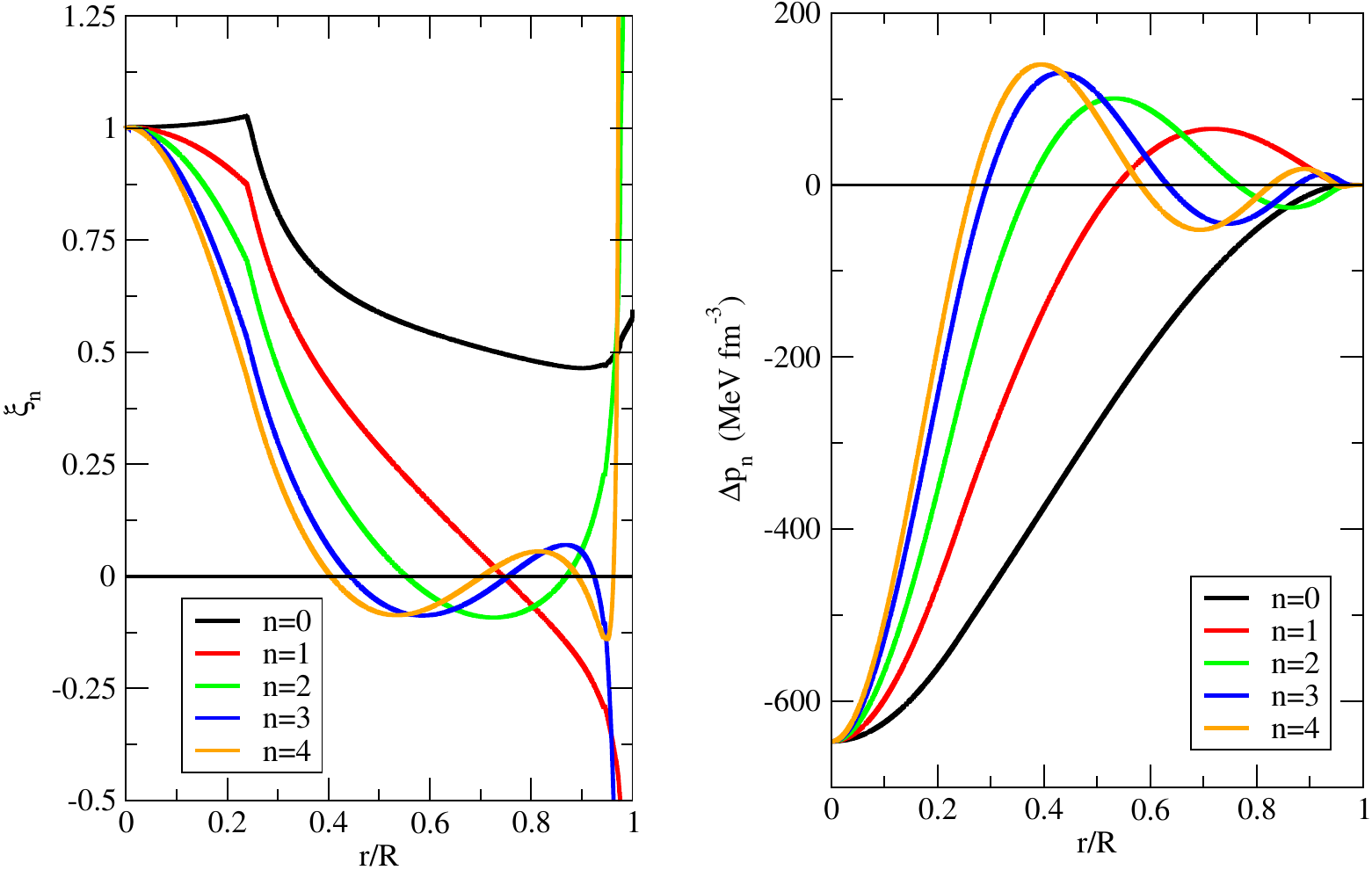}
    \caption{Same as Fig. \ref{fig:xinpe} of a  $2M_\odot$ star for antikaon optical potential $-120$ MeV.}
    \label{fig:xi120}
\end{figure}

\begin{figure}
    \centering
    \includegraphics[scale=.30]{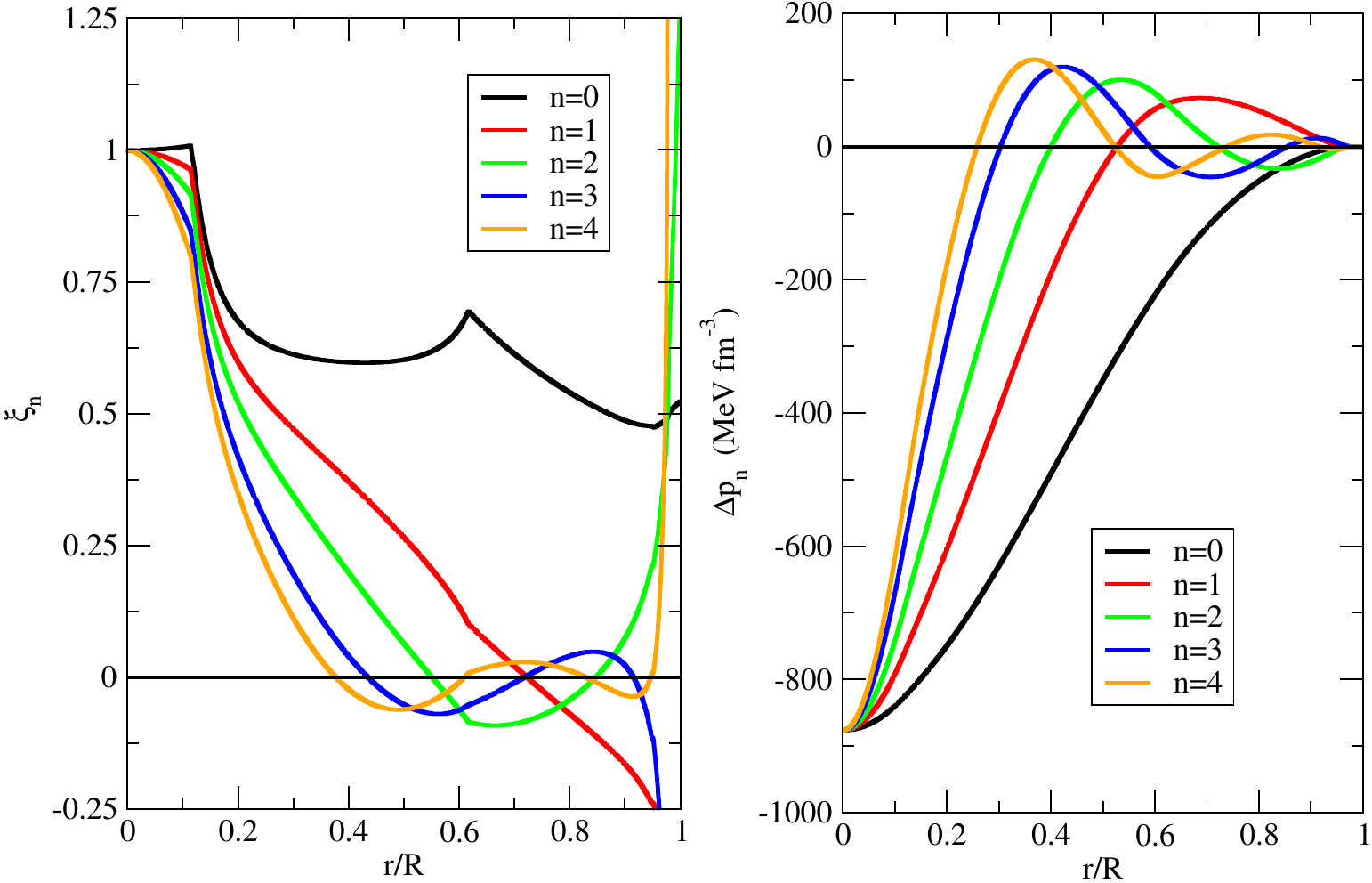}
    \caption{Same as Fig. \ref{fig:xinpe} of a  $2M_\odot$ star for antikaon optical potential $-160$ MeV. }
    \label{fig:xi160}
\end{figure}

In Fig.~\ref{fig:xinpe}, the eigenfunctions, $\xi_n$ and $\Delta P_n$ for an NS made of nucleon-only matter are shown (with $M=2.0 M_\odot$ and $R =13.10$ km). These eigenfunctions are computed for the low-order radial modes, $n = 0, 1, 2, 3, 4$, shown in black, red, green, blue and orange, respectively. Similarly, in Fig.~\ref{fig:xi120}, the $\xi_n$ and $\Delta P_n$ eigenfunctions are shown for the NS made of antikaon condensates for $U_{\bar K}= -120$ MeV  (with $M=2.0 M_\odot$ and $R =13.11$ km) and in Fig.~\ref{fig:xi160}, the same for $U_{\bar K} = -160$ MeV  (with $M=2.0 M_\odot$ and $R =12.43$ km). So from these three figures, it is clearly seen that the antikaon condensation in the nuclear system affects the eigenfunction  $\xi_n$. Two distinct kinks appear in the eigenfunctions $\xi_n$ for the onset of antikaon in the system and the kinks are more intense in the case of optical potential $-160$ MeV. But, for the eigenfunctions $\Delta P_n$, antikaon condensation does not affect at all. The differences between the amplitude of successive eigenfunctions $\xi_{n+1}-\xi_n$ are maximum at the star surface, and they also acquire an opposite sign.

\section{Summary and Conclusions}
\label{summary}

In this work, we have studied the structure and the composition of NSs with antikaon condensates. We have used a RMF model with density dependent coupling for the nucleons. However, the antikaon-nucleon couplings are not considered as density dependent. We have studied different values of the antikaon optical potentials to study antikaon condensed phase. For all the cases, we have found that the antikaons appear in the system as a second order phase transition. But, there is a caveat concerning the appearances of the $\bar{K^0}$. The off-shell condition for the $KN$-scattering amplitude suggests that the interaction becomes repulsive at $\omega_K \rightarrow 0$, according to the low energy theorem \cite{Yabu:1993jk,Thorsson:1995rj}. Incidentally, this is also the threshold condition for $\bar{K^0}$. Hence, there is an ambiguity whether it will appear in the system or not. In this work, we have considered only the on-shell conditions for the antikaons following Pal et al. \cite{Pal:2000pb}, and did not include the off-shell corrections which is beyond the scope of this work. Then, we continue to calculate the structure of stars for different possible EOSs and find that very deep potential ($\lesssim -170$ MeV) leads to a maximum mass less the $2M_\odot$. Then we have studied the radial oscillations for all stellar configurations of the mass-radius sequence. We have also calculated the eigenfunctions corresponding to a $2M_\odot$ star for $U_K = -120, -160$ MeV to compare the effect of the phase transition on the fundamental and the overtone frequencies and eigenfunctions. We have found that the effect is most prominent on the fundamental eigenfunctions but not as much on the higher order eigenfunctions. In case of the overtone frequencies, they are more affected than the fundamental frequencies. In this context, we would also like to mention that we have found two approaches for the solutions of the oscillations in the literature. One is by Vaeth and Chanmugam \cite{1992A&A...260..250V} and the other by Gondek et al. \cite{Gondek:1997fd} differing on how to define the pressure perturbation. We have noticed the most of the works following the first approach do not show the profiles of different quantities of interest upto the surface. We speculate the reason to be due to the numerical issues resulting the term $\Delta P/P$ blowing up near the surface. While the approach by Gondek et el. does not produce such irregularities. This issue has also been discussed in Pereira et al. \cite{Pereira:2017rmp}. Our results fully match with their assertion on this point. Our future aim would be to extend this study for the PNSs. We will also investigate the effects of different damping mechanisms that maybe important in a dynamical scenario. We will also check the effects the appearance of $\Lambda$ hyperons in addition to the antikaons.

\section*{Acknowledgements}
We thank the reviewers for their useful comments that helped us improving the manuscript. We also thank Professor D. Bandyopadhyay for many useful discussions. P. Char is currently supported by the Fonds de la Recherche Scientifique-FNRS, Belgium, under grant No. 4.4501.19.

\bibliographystyle{epj}
\bibliography{mybiblio}
\end{document}